# Towards Near Real-Time BGP Deep Analysis: A Big-Data Approach

Joel Obstfeld
Cisco
UK
jobstfel@cisco.com

Xiaoyu Chen
Cisco
UK
williscc@cisco.com

Olivier Frebourg
Cisco
UK
ofrebour@cisco.com

Pavan Sudheendra
Cisco
UK
pasudhee@cisco.com

## ABSTRACT

BGP (Border Gateway Protocol) serves as the primary routing protocol for the Internet, enabling Autonomous Systems (individual network operators) to exchange network reachability information. Alongside significant on-going research and development efforts, there is a practical need to understand the nature of events that occur on the Internet.

Network operators are acutely aware of security-related incidents such as 'Prefix Hijacking' as well as the impact of network instabilities that ripple through the Internet. Recent research focused on the study of BGP anomalies (both network/prefix instability and security-related incidents) has been based on the analysis of historical logs. Further analysis to understand the nature of these anomalous events is not always sufficient to be able to differentiate malicious activities, such as prefix- or sub-prefix- hijacking, from those events caused by inadvertent misconfigurations. In addition, such techniques are challenged by a lack of sufficient resources to store and process data feeds in real-time from multiple BGP Vantage Points (VPs).

In this paper, we present a BGP Deep-analysis application developed using the PNDA (Platform for Network Data Analytics) 'Big-Data' platform. PNDA provides a highly scalable environment that enables the ingestion and processing of 'live' BGP feeds from many vantage points in a schema-agnostic manner. The Apache Spark-based application, in conjunction with PNDA's distributed processing capabilities, is able to perform high-level insights as well as near-to-real-time statistical analysis.

## 1 INTRODUCTION

On the 26[th] of April, 2017, a 'prefix hijacking' event occurred that affected a number of financial services companies around the world. The impact of the event was such that traffic was in part, diverted and directed to another network that claimed to be the owner of the IPv4 address space.

The nature of the event was such that, depending on 'distance' between your network and that of the impacted companies, the 'newly' announced network would now appeared be 'closer' (or 'shorter' in BGP terms) and therefore would be preferred.

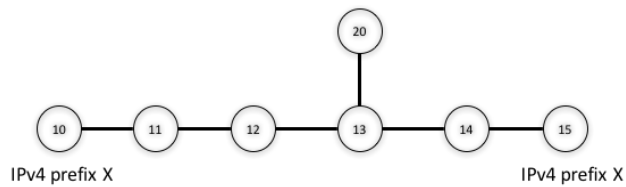

Figure 1: BGP AS Path length preference

In the diagram above, the network operator responsible for Autonomous System (AS) 10 advertises the IPv4 prefix X. From the position of AS20, the 'distance' is considered to be an AS Path length of 4, (AS13, AS12, AS11, AS10). In the case of prefix hijacking, the same IPv4 prefix, X, was announced as being located in AS15. Given the proximity of AS20 to AS15 with an AS Path length of 3, traffic that originally flowed from AS20 toward AS10 will now be directed to AS15. However, if you were located in AS11 or AS12, the AS Path length is still shorter towards AS10 than towards AS15 and as a result, traffic will continue to flow towards AS10.

An additional element of the 26[th] of April event, was the announcement of 'more-specific prefixes' from the source of the attack, also known as 'sub-prefix injection'. The BGP protocol's path selection algorithm prefers 'more-specific' prefixes before considering the AS Path length.

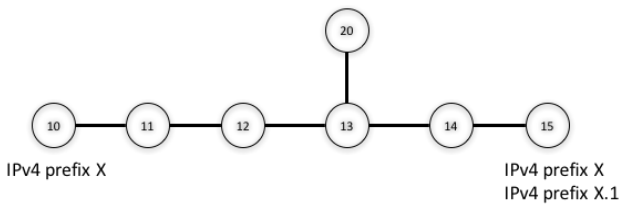

**Figure 2: BGP AS Path length with 'longer prefix'**

In the diagram above both AS10 and AS15 announce the IPv4 prefix X. In addition, AS15 announced the 'more-specific prefix', X.1. Since routers in the Internet operate on the premise of 'longest match', in such a case, regardless of the AS path length, the path to the more-specific prefix (X.1) will be preferred and traffic from all AS's other than AS10, destined to addresses in the X.1 address range, will be diverted towards the attacker's network (AS15).

For the purposes of illustration, let us assume that AS10 announced the IPv4 prefix 123.123.0.0/18. During the attack, AS15 also announced the IPv4 prefix 123.123.0.0/18 and in addition, the IPv4 prefix 123.123.63.0/24. Given the use of 'longest match', any traffic destined to an IPv4 address in the range 123.123.63.0-255 would be sent towards AS15, rather than the legitimate origin AS.

While it is possible that the announcement was as a result of an error, the announcement of more-specific prefixes that 'targeted' the financial institutions, makes it less likely to have been a mistake. There are well-known solutions such as RPKI[1], that provide a mechanism by which one can verify the authenticity of prefix announcements but gaps in adoption and usage of the RPKI solution provide the opportunity for further events to occur.

It should also be noted that if network operators follow BGP security best practices, in respect to filtering incoming advertisements and specifically blocking the receipt of their own prefixes, they are typically unaware of a prefix hijack incident taking place until an impacted user contacts them!

In order to be able to detect and mitigate a prefix hijack, network operators need to have an ability to 'see' outside of their network perimeter, in order to understand what other networks 'see' in respect to the prefixes relating to that AS.

With at the time of writing, over 660,000 prefixes present in the Internet routing table, looking for erroneous BGP events such as the one described above, is akin to looking for a 'needle in a haystack'. In a world where real-time system, such as those used by the financial services industry, are impacted, real-time, or close to real-time detection and reporting system are also required.

In this paper, we will describe how event analysis can be performed over the BGP data-set using Big-data analysis techniques, and that we can move towards a position where real-time events can be detected and reported in real-time. In addition, we will describe a generic Big-data platform that can be readily adapted to consume and process a wide range of data types.

## 2 Related Work

According to the recent technical review [2], techniques in BGP anomaly detection have been dominated by time-series analysis, machine learning, statistical pattern recognition, and validation of BGP updates based on history log. These significant works are either limited to identifying some anomalies or focused on 'after-event' analysis based on datasets where available. As the survey correctly pointed out in its' conclusion, there is still much to be done to detect BGP anomaly in real-time such that network operators can mitigate the spreading of anomalies across networks.

Due to the 'openness' of the BGP protocol, one can claim the false ownership of one or more prefixes (i.e. prefix hijacking) and start a Denial-of-Service (DoS) attack. Such false prefix path announcement can also be propagated over networks due to policy misconfigurations. Existing automated anomaly detection techniques with rules learned from a 'stable' environment (experimental testbed or historical data) can easily become obsolete overtime and fail to deal with modern network attacks.

In parallel to automated anomaly detection techniques, research efforts have also focused on advanced visualization techniques to help network operators to understand BGP routing instability and identify potential anomalies. Such visualization tools range from those that visualize raw BGP update events, such as BGPlay[3] and LinkRank[4], to those with additional statistical view to show historical trends and current deviations, such as BGP eye[5] and Vistracer[6].

The emergence of big data technologies makes it possible to combine machine-automated analysis and exploratory analysis in a hybrid manner, in order to provide anomaly detection in real-time or near real-time. Machine-automated processes are used to generate alerts based on patterns/deviations learned from incrementally increasing historical data and live updates. Visualization tools then helps network operators to dive into observed BGP events and validate reported alerts. In 2016, the Center for Applied Internet Data Analysis (CAIDA) published BGPStream [6], an open-source software framework for both offline analysis of historical MRT feeds and real-time prefix/AS monitoring. There are also commercial players Kentik [8] and BGPmon (OpenDNS) [9] that provide platforms for network data analytics and real-time intelligence.

The work presented in this paper uses a similar semi-automated approach, leveraging an open-source BGP feed collection framework in conjunction with an open-source big data platform. Compared to other existed approaches, we provided a versatile and scalable end-to-end solution that enables the collection of live BGP updates in a schema-agnostic manner. In addition, the work offers the ability to process data at scale as well as being a platform that can be applied to other data-sets and applications.

## 3   Goals and Challenges

BGPv4 is the control-plane protocol that drives the Internet. The Internet itself is a highly volatile environment; from a single vantage point, we observe on average ~ 3 million BGP updates a day. Our main goal was to create a solution to help network operators extract useful signals from the 'noise' such that they can understand the changes in the Internet that may impact their own operations, and to do so in near real-time.

In order to gain better insight, it is important to perform the analysis on data taken from multiple BGP Vantage Points (VPs), each of which has a view of the Internet. Collecting BGP data from globally distributed VPs poses a series of technical difficulties. Due to the distributed nature of VPs, BGP event data cannot be readily synchronized.

How one can collect BGP updates from the VPs in a time-synchronous and format agnostic manner, becomes the first technical issue. In this work, we leverage an open-source project, SNAS.io [10](Streaming Network Analysis System), that supports streaming live BGP updates and is highly scalable.

Another challenge comes from the accumulation of data over time. There are two issues related to this. First, we need to provide a scalable storage. Secondly, we need a solution that will enable us to perform the required analysis in near real-time. In this work, we adopted state-of-the-art big data technologies and leveraged the PNDA.io [11], an open-source big data framework. With PNDA, we were able to utilize a file system distributed across multiple data nodes to store historical data meanwhile minimizing processing time in batch mode to keep pace with live data streaming.

Finally, the solution aims to provide a versatile and extensible solution for a broad range of applications and users. Using advanced visualization techniques, we are able to provide continuous monitoring on BGP anomalies along with capability of investigating historical trends from three different viewpoints, AS connectivity, AS path view, and prefix view.

## 4   End-to-End Architecture

Figure 3 shows the end-to-end architecture of BGP deep analysis application, which consists of three main layers:

- Data collection layer
- Data storage & processing layer
- Data visualization layer

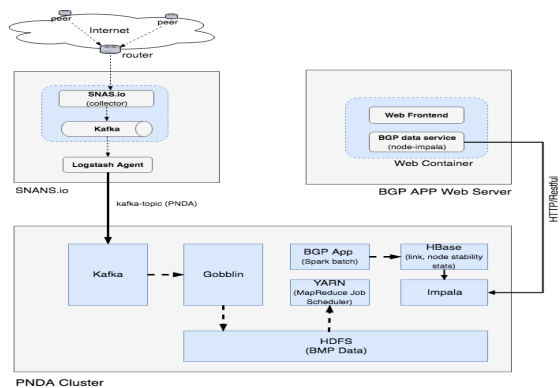

**Figure 3: BGP deep analysis end-to-end technical architecture**

## 4.1 Data collection and aggregation

In order to ingest BGP information, the system makes use of an open-source project called SNAS.io. SNAS supports the consumption of data presented in the BGP Monitoring Protocol (BMP) format [12].

In many cases, BGP monitoring is performed using an instrumented BGP 'listener' process running on a compute host. The process is added into a network operator's BGP topology and therefore hears the BGP event updates that pass across the network. A major limitation of this method is that it relies on BGP's advertisement selection process. The selection process will consume all of the received BGP updates, make its 'best-path' selection and advertise only the best-paths (Post-Rib). Those received paths that have not been selected are discarded and more importantly in the context of this work, cannot be readily obtained from the router.

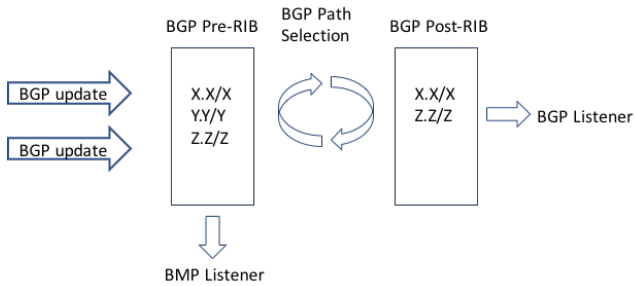

**Figure 4: BGP data source**

Routers that support the BMP protocol are able to send to a BMP 'listener', such as SNAS.io, all of the BGP updates that router has received before the BGP path selection algorithm comes into effect (Pre-Rib). This key capability means that the system can record events before any form of filtering takes place. In addition, SNAS.io supports RouteViews [13], enabling the system to obtain observations from many locations around the world.

Once the data has been received by SNAS.io, the data is passed to a Logstash [14] process, an open-source component that is used to consume and in this case, encode the BMP data for storage in the PNDA platform's Hadoop [15] storage system. The BMP data is encoded in an avro [16] wrapper and placed onto the PNDA system's Kakfa bus. The wrapper provides common metadata that is used by Gobblin [17] in order to organize and write the data within the HDFS folder on a per-topic, per-hour, per-BMP 'listener' basis. The BGP data is now available for applications to inspect and process.

One should note that a wide variety of data formats can be processed and encoded using Logstash or could be written directly to the Kafka [18] bus. This makes the PNDA system flexible and extensible.

## 4.2 Underlying big data platform

Consuming and processing data of a particular type and construct can be performed using specifically-optimized processing pipelines. Such pipelines are typically designed for deployment in a specific environment, reducing the ability to reuse the solution in other environments or with other data types. While there may be performance optimizations that can be realized as a result of the design and technology selection, one is left with a bespoke solution.

When considering the technologies that can be applied to the particular problem space, it is valuable to look at adjacent fields for inspiration and for a sense of 'scale'. What may be considered to be a sizeable data-set in one field of work, is considered small or even insignificant in another.

The BGP Deep Analysis work is considered to be an application running on top of generalized 'Big-data' platform. For many people, 'Big-data' is synonymous with 'Hadoop' and a series of technologies that have grown up in this space. While the use of Hadoop ecosystem technologies is common-place in some areas of business and technology today, they have not been applied to the networking space until very recently.

Moreover, building a Big-data platform does not stop with the deployment of a Hadoop cluster. Additional components are required in order to be able to ingest data, potentially in a wide variety of forms, volumes and velocities. Another set of components are required to expose the data in a manner that Data Scientists can interact with. The rate of developments in open source Big data analytics technologies is extremely rapid, but combining multiple technologies into an end-to-end solution can be extremely complex and time-consuming.

The current state of the networking industry is that each network operator is solving this problem independently, leveraging upstream Hadoop distributions to build custom analytics platforms and applications. These custom implementations become operator-specific siloes, which are not interoperable; each operator owns the software lifecycle and associated development costs for their implementation.

For the BGP Deep Analysis work, the open source PNDA environment has been used. PNDA brings together a number of open source technologies to provide a scalable open platform on top of which analytics applications can be developed and deployed. PNDA provides a Hadoop cluster, with data distributed across 'data-nodes' both for efficiency of processing and for resilience, as well as the components and capabilities required to offer support for multiple end-to-end pipelines, on a single platform.

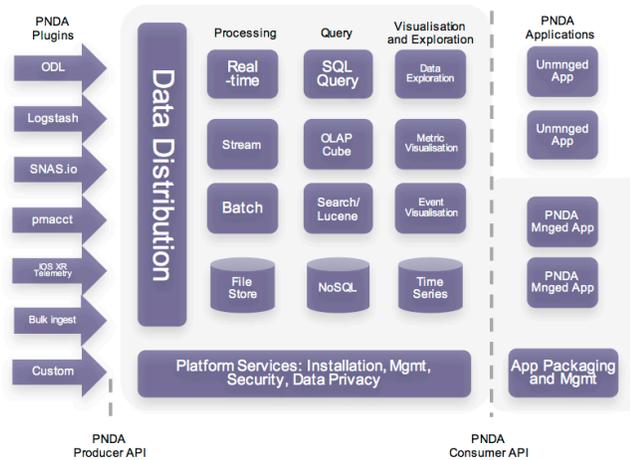

**Figure 5: PNDA block diagram**

A fundamental premise of 'Big-Data' is to obtain and store data in its rawest form. In many traditional data applications, data is modified or transformed on ingest, resulting in the reduction of information as the data is 'massaged' into a required format. A side-effect of this method of operation is that if another application wishes to make use of the data, it is already operating on a refined subset, rather than the initial 'raw' information.

The PNDA platform supports a wide-range of off-the-shelf plugins as well the creation of custom adapters, enabling the ingestion of structured and unstructured data such as log messages, SNMP events, metrics/time-series data and network telemetry.

Another premise of a Big-data platform is the performance of processing where the data resides, rather than moving the data from storage to a location where the analysis application executes. The platform supports the use of Apache Spark-based batch and streaming applications

### 4.3 BGP app

Similar to conventional three-tier Web application, the BGP Web server presented a Web-based UI presenting to the operator results generated by two Spark-based applications, the BGP deep analysis application and the BGP security application. Both applicaitions are scheduled as Apache Spark micro-batch processes and process raw BGP data streamed into PNDA via Kafka brokers and then persisted in HDFS by a Gobblin process.

The BGP deep analysis application is scheduled on frequent interval basis and processes raw BGP data to aggregate behaviors on a per-AS and per-AS Path basis. In particular, it characterizes the stabilities of AS and AS path using the key features as listed in Table 1. The key features obtained from the BGP data are the 'AS path change' (announcement and withdrawal of a prefix), the rank value derived from the number of prefixes that transit via a particular AS or a particular inter-AS path, rank change, and rank change frequency for each AS and AS path. By using these features we can derive the importance of an AS and AS path over the Internet as well as the stability (the higher change frequencies, the less stable).

| ID | Feature |
|---|---|
| 1 | AS Path change |
| 2 | AS rank |
| 3 | AS path rank |
| 4 | AS rank change |
| 5 | AS path rank change |
| 6 | AS rank change frequency |
| 7 | AS path rank change frequency |

**Table 1: Key fields for AS analysis**

In parallel to the BGP deep analysis application, the BGP security application is scheduled to analyze network anomalies including RFC5735 [19] prefixes and netblocks that are unallocated and therefore should not ever seen in BGP updates, short prefixes (i.e. those prefixes with a mask between /0-/7 and those with a mask of /25-/32. Further work identifies those prefixes that originate from multiple AS's and 'sub-prefix injection' cases. Table 2 lists the key features used to detect network anomalies.

| ID | Feature |
|---|---|
| 1 | Prefix origin change |
| 2 | Prefix length |
| 3 | Rare Prefix |
| 4 | Subprefix origin change |

**Table 2: Key fields for Security analysis**

Both applications are running in batch mode and schedule apache Spark MapR jobs across data nodes via Hadoop YARN resource manager, in short 'spark-on-yarn'. At the end of batch processing, the applications write their results into HBase tables, a key-value database, and make them available via an Impala query interface. The front-end portal provides a Web-based UI that notifies network operators to network anomalies and employs advanced visualization techniques to enrich the network operators' understandings of the nature of anomalies, as well as enabling the investigation of data using long historical analysis.

## 5  BGP Deep Analysis

The Deep Analysis applications investigate two key areas today; BGP AS path and prefix behavior and BGP Security. With the data platform consuming event data from multiple VPs, the application is able to process aggregate data to understand, for example, which AS's appear in AS Paths most frequently? Which AS's originate the largest number of prefixes? Which AS's exhibit the largest number of 'changes' over time.

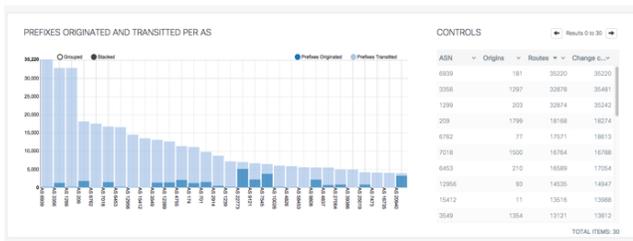

**Figure 6: Top N analysis**

Using further mapping techniques, we can determine the connectivity characteristics of a particular AS, rendering the results in a graph.

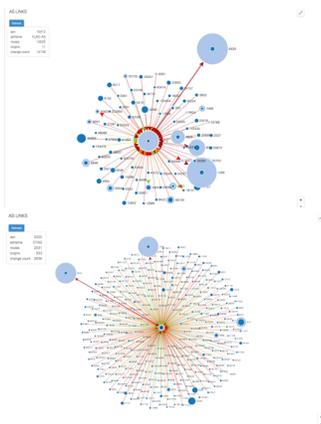

**Figure 7: Plot comparison to two AS's**

Focused analysis on paths between two AS has identified in one 24-hour period:

- Shortest path – 4 hops
- Longest path – 29 hops
- Longest unique AS path – 6
- Unique paths - 9
- Largest prepend count – 17
- Prepend variation – [7-17]

The Security analysis looked for BGP prefixes that should not be observed. This includes looking for 'unallocated' address space. Observations over a 12-hour period revealed over 4000 unique 'unallocated' prefixes being announced. In addition, a module was developed in order to identify cases of 'sub-prefix injection'. At the time of submission, this study was still ongoing.

## 7  Conclusions

The work covered in this paper has shown how techniques and technologies from the Big-data realm can be applied to the Networking space. The work has demonstrated the use of applications that examine, identify and alert network operators to events that have the potential to be highly disruptive to their customers and services.

In addition, the use of a Big-data platform enables the further exploration and analysis of other data sets both inside and outside of the networking realm, without having to create entirely new data acquisition, storage and processing pipelines.

## ACKNOWLEDGMENTS

This work was realized with the assistance of the PNDA and SNAS project teams at Cisco. PNDA.io and SNAS.io are open source projects under the Linux Foundation.

## APPENDIX

### Appendix 1

The figure below shows a simplistic graph. The 'top' line shows the original prefix announcement, originating from AS 13118. On the 26[th] of April, at 22:40, the 'sub-prefix hijack' was advertised, originating in AS12389. This is shown in the 'spot' point in the lower third of the graph. This announcement led to significant service interruption for the affect institution.

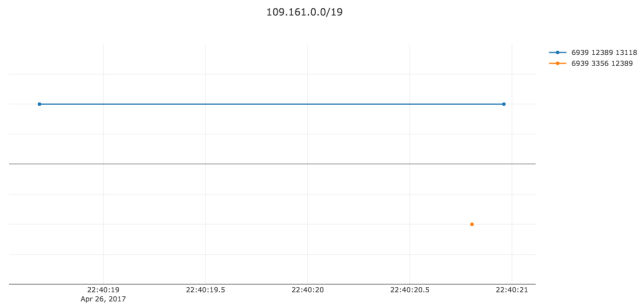

**Figure 1: Prefix Hijacking of attacking financial institutions, Apr. 26th, 2017**

## REFERENCES
[1] R. Bush, R. Austein, https://tools.ietf.org/html/rfc6810, 2013
[2] B. Al-Musawi, P. Branch and G. Armitage, "BGP Anomaly Detection Techniques: A Survey," in *IEEE Communications Surveys & Tutorials*, vol. 19, no. 1, pp. 377-396, Firstquarter 2017. doi: 10.1109/COMST.2016.2622240
[3] Di Battista G., Mariani F., Patrignani M., Pizzonia M. (2004) BGPlay: A System for Visualizing the Interdomain Routing Evolution. In: Liotta G. (eds) Graph Drawing. GD 2003. Lecture Notes in Computer Science, vol 2912. Springer, Berlin, Heidelberg
[4] M. Lad, D. Massey, and L.Zhang. Linkrank: A graphical tool for capturing bgp routing dynamics. In Proceedings of the IEEE/IPIF Network Operations and Management Symposium (NOMS), April 2004
[5] S. T. Teoh, S. Ranjan, A. Nucci, and C.-N. Chuah. Bgp eye: a new visualization tool for real-time detection and analysis of bgp anomalies. In VizSEC '06: Proceedings of the 3rd international workshop on Visualization for computer security, pages 81–90, New York, NY, USA, 2006. ACM.
[6] Asdf Orsini, C., King, A., Giordano, D., Giotsas, V. and Dainotti, A., 2016, November. BGPStream: a software framework for live and historical BGP data analysis. In *Proceedings of the 2016 ACM on Internet Measurement Conference* (pp. 429-444). ACM.
[7] Fischer, F., Fuchs, J., Vervier, P.A., Mansmann, F. and Thonnard, O., 2012, October. Vistracer: a visual analytics tool to investigate routing anomalies in traceroutes. In *Proceedings of the ninth international symposium on visualization for cyber security* (pp. 80-87). ACM.
[8] https://www.kentik.com/
[9] https://bgpmon.net/
[10] http://www.snas.io/
[11] http://www.pnda.io/
[12] https://tools.ietf.org/html/rfc7854, 2016, Scudder, Fernando, Stuart
[13] http://www.routeviews.org/
[14] https://www.elastic.co/products/logstash
[15] http://hadoop.apache.org/
[16] https://avro.apache.org/
[17] https://engineering.linkedin.com/data-ingestion/gobblin-big-data-ease
[18] https://kafka.apache.org/
[19] M. Cotton, L. Vegoda, https://tools.ietf.org/html/rfc6598, 2010